\documentclass{aa}
\usepackage{graphicx}
\begin{document}
\title{NGC 146: A young open cluster with a Herbig Be star and intermediate mass pre-main sequence stars}

\author{A. Subramaniam\inst{1} \and D.K. Sahu\inst{2} 
        \and R. Sagar\inst{3} \and P. Vijitha\inst{4}}
\offprints{Annapurni Subramaniam}
\institute{Indian Institute of Astrophysics, II Block Koramangala, Bangalore 560 034\\
          \email{purni@iiap.res.in}
        \and
            CREST campus, Indian Institute of Astrophysics, Hosakote, Bangalore,\\
           \email{dks@crest.ernet.in}
        \and
         Aryabhatta Research Institute of Observational Studies (ARIES),\\
          Manora peak, Nainital - 263129, India\\
         \email{sagar@upso.ernet.in}
         \and
         VSRTP student, Mother Theresa University, Kodaikanal\\
         }
\date{Received / accepted}
\titlerunning{NGC 146: A young open cluster with a Herbig Be star}
\authorrunning{Subramaniam et al.}

\abstract{We present UBV CCD photometry and low-resolution spectra of stars in the
field of the young open cluster NGC 146. UBV photometry of 434 stars were used to
estimate the E(B$-$V) reddening of 0.55 $\pm$ 0.04 mag 
and BV photometry of 976 stars were used to estimate a distance modulus of
(m$-$M)$_0$ = 12.7 $\pm$0.2 mag, corresponding to a distance of
3470$^{+335}_{-305}$ pc. We estimated 10 -- 16 Myr as the turn-off age for the upper
main sequence of the cluster using isochrones and synthetic colour magnitude diagrams.
We identified two B type stars with H$_\alpha$ in emission and
located on the MS using slit-less spectra. A higher resolution spectrum of the brighter Be star indicated the
presence of a number of emission lines, with some lines showing the signature of gas infall. This star was found 
to be located in the region of Herbig Ae/Be stars in the
(J$-$H) vs (H$-$K) colour-colour diagram. Thus, we identify this star as a Herbig Be star. 
On the other hand, 54 stars 
were found to show near infrared excess, of which 17 were found to be located
in the region of Herbig Ae/Be stars and 18 stars were found to be located in the region of Be stars in
the NIR colour-colour diagram. Thus NGC 146 is a young cluster with a large number of
intermediate mass pre-main sequence stars. The turn-on age of the cluster is found to be $\sim$ 3 Myr. 
Though NGC 146 shows an older turn off, 
the bulk of stars in this cluster seems to belong to the younger population of 3 Myr.

\keywords{Galaxy:Open cluster:NGC 146, Stars: Be, Stars: CM diagrams, Stars: pre-main sequence}
}
\maketitle
\section{Introduction}
The open star cluster system of our Galaxy is one of the important constituent
of the disk of the Galaxy. The open star clusters are formed from molecular clouds
at the sites of star formation.  NGC 146 is a young open cluster
located in the direction of the Perseus spiral arm.
NGC 146 (RA = 0h 33.05m; Dec = +63 18.1 (J2000)) is located away from 
the center of the Galaxy (l=120.9; b=0.5). The approximate age of the
cluster was estimated as 10 Myr and the distance to the cluster was
found to be more than 3 kpc, which reinforces the assertion that there may be
a spiral arm beyond the Perseus arm (Phelps \& Janes \cite{pj94}). 
In the present study, we have obtained UBV photometry 
of the stars in the field of NGC 146 up to a limiting magnitude of V $\sim$ 20 mag.
This has helped in estimating the 
differential reddening across the field of the cluster, thereby obtaining better
estimates of the cluster parameters.  The cluster has a number
of early B type main-sequence stars indicating its young nature, but there are 
no super giants.
The absence of evolved stars has hampered the estimation of age of the cluster
using the isochrone fitting method. In this study, we estimate the age of the cluster
using the upper main-sequence (MS) as well as pre-main sequence (PMS) stars.

This cluster has been studied previously by Hardorp (\cite{h60}) using RGU photometry and 
Jasevicius (1964) using UBV photographic photometry. Phelps \& Janes (\cite{pj94}) obtained
CCD photometry of this cluster in UBV passbands. The limit of the photometry was
V $\sim$ 18.0 mag. They found the presence of variable reddening across
the field of the cluster. The mean reddening was found to be E(B$-$V) = 0.70 mag.
They estimated a distance of 4786 pc and found that the cluster is younger than 10 Myr.

\section{Observation and data analysis}
The cluster was observed using the 2.0 m Himalayan Chandra Telescope using the HFOSC instrument, in the
imaging mode, on 3 July 2003. The CCD used for imaging is a 2 K $\times$ 4 K CCD,
where the central 2 K $\times$ 2K pixels were used. The pixel size is 15 $\mu$
with an image scale of 0.297 arcsec/pixel. The total area observed is 
approximately 10 $\times$ 10 arcmin$^2$. The log of the observations is tabulated in 
Table 1.

\begin{figure}
\resizebox{\hsize}{!}{\includegraphics{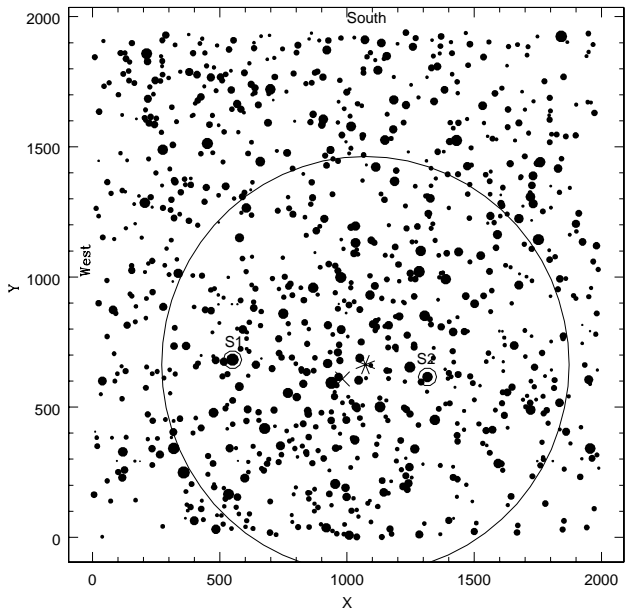}}
\caption{The observed region of the cluster NGC 146 is shown here. S1 and S2
denote the stars which show H$_\alpha$ line
in emission. The estimated center of the cluster is marked by asterisk sign. The
radius where increament in the stellar density as seen in the RDP is shown by the
big circle. }
\label{figure1}
\end{figure}
The night of observation was not photometric and hence
we used 18 bright stars observed by Phelps \& Janes (\cite{pj94}) as calibration stars.
The CCD data were calibrated using IRAF data reduction package. The 
instrumental magnitudes were estimated using the DAOPHOT II package.
The zero point errors are of the order of 0.03 mag. The V magnitude and
(B$-$V), (U$-$B) colours were estimated for 483 stars, whereas only
V magnitude and (B$-$V) colour were estimated for 976 stars. The observed
field of the cluster with 976 stars is shown in figure 1.
From the above data, only stars with errors less than 0.1 mag is used for
the estimation of cluster parameters.
The cluster was also observed in the slit-less mode with the grism as the dispersing 
element using the HFOSC. This mode of observation using the HFOSC yields an image, where
the stars are replaced by their spectra. This is similar to objective prism spectra.
The broad band R
filter (7100\AA,BW=2200\AA) and Grism 5 (5200-10300\AA) of HFOSC CCD system was used in
combination without any slit. The resolution of grism 5 is 870.
We used the Johnson's R filter in the field, to restrict the spectra to the 
spectral region of the R band. The slit-less spectra thus obtained
were used to identify the presence of any 
emission line stars in the field of the cluster. Since NGC 146 is known to be a young 
cluster with a number of stars populating the B-spectral range, there is a good chance
of some of the above stars showing emission line feature.
The first two slit-less spectra were obtained on 27 January 2004.
We first obtained an image of the cluster in R filter, followed by 1 and 10 minute
exposures with the grism. The image is shown in figure 2.
The observations were repeated on 25 June, 21 July and 23 November 2004.
Low resolution spectra of two stars were obtained on 18 January 2005 in order
to study the spectral features in detail.
\begin{figure}
\resizebox{\hsize}{!}{\includegraphics{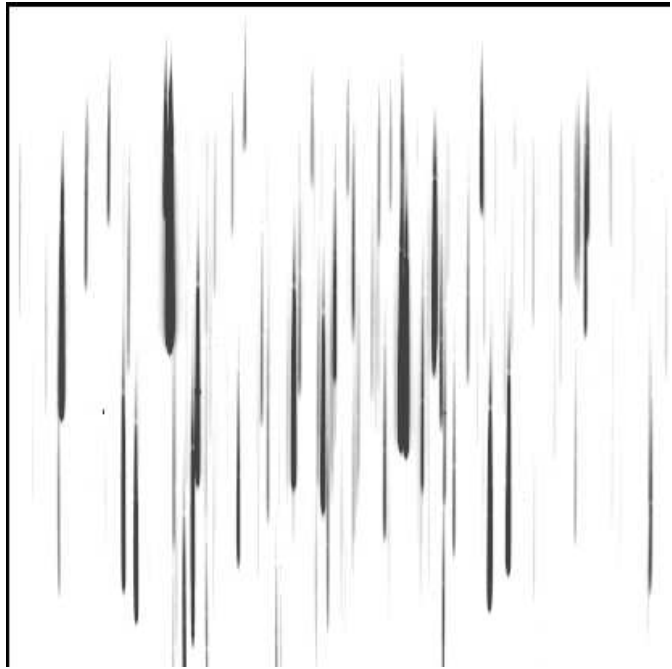}}
\caption{The slit-less spectra of stars in the region of NGC 146 is shown.
This is obtained by using an R filter and a grism in combination. This image is
a co-added one, with total exposure time of 30 min. The H$_\alpha$ emission in two
spectra can be noticed. The one on the left is S1 and that on the right is S2.}
\label{figure2}
\end{figure}

\begin{table*}
\caption{Journal of observation.}
\begin{tabular}{ccccc}
\hline
 Date    &    Filter&  UT &   Airmass& Exp. Time (Sec)\\
\hline
Imaging & & & & \\
July 03 2003 &   V &  21 22& 1.36&  10.0 \\ 
July 03 2003 &   V &  21 26& 1.35&  60.0 \\ 
July 03 2003 &   V &  21 30& 1.34&  60.0 \\ 
July 03 2003 &   V &  21 34& 1.33&  60.0 \\ 
July 03 2003 &   V &  21 40& 1.32&  60.0 \\ 
July 03 2003 &   U &  21 54& 1.29& 120.0 \\ 
July 03 2003 &   U &  21 59& 1.28& 120.0 \\ 
July 03 2003 &   B &  22 13& 1.26& 300.0 \\ 
July 03 2003 &   B &  22 21& 1.25& 300.0 \\ 
July 03 2003 &   B &  22 30& 1.24&  90.0 \\ 
Slit-less spectra& & & & \\
27 January 2004 & R + Grism 5 &  & --& 60.0 \\
27 January 2004 & R + Grism 5 &  & --& 600.0 \\
25 June 2004& R + Grism 5 &  & --& 900.0 (2 exposures) \\
21 July 2004& R + Grism 5 &  & --& 900.0 (2 exposures) \\
23 November 2004& R + Grism 5 &  & --& 900.0 (1 exposure) \\
Slit spectra (S1)& & & & \\
18 January 2005 & Gr7 & & --&  900.0 (3500 -- 7000 \AA)\\
18 January 2005 & Gr8 & & --&  600.0 (5200 -- 9200 \AA) \\
Slit spectra (star 1)& & & & \\
18 January 2005 & Gr7 & & --&  600.0 (3500 -- 7000 \AA)\\
18 January 2005 & Gr8 & & --&  600.0 (5200 -- 9200 \AA) \\
\hline
\end{tabular}
\end{table*}

\section{Structure of the cluster}
From the cluster plot as  shown in figure 1, it can be seen that
the cluster does not have a well defined core. Hence the center of the cluster
is not well defined from the visual inspection. The cluster center was estimated by
taking the average value of the X and Y coordinates
of all the stars within a radius of 700 pixels, after initially adopting
a visual center. All stars brighter than V = 19 mag
were considered for the center estimation.
The estimated center thus found is marked by a cross in figure 1. 
We estimated the radial density
profile (RDP) to study the radial structure of the cluster.
The stellar density are estimated in radial bins from the center.
We computed the RDP for radial bins of 0.5 arcmin. RDP were estimated for
stars brighter than V = 19 mag and also for stars brighter than 17 mag. 
The RDP estimated by considering stars brighter than V = 19 mag
was found to have a lot of scatter, on the other hand, a much better and 
smoother profile was obtained when stars brighter than 17 mag was considered.
The cluster center estimated using only these bright stars is shown as asterisk in figure 1.
It can be seen that the two centers are slightly different.
This RDP is shown in figure 3. The dots which are connected by the solid line 
represents the RDP. It can be seen that within a radius of 4 arcmin, the profile
is more or less smooth. Beyond 4 arcmin radius, the profile shows an increament in
the stellar density, instead of a decreament. This behaviour is also seen in the 
profile estimated by considering more fainter stars, but not as prominent as seen here.

\begin{figure}
\resizebox{\hsize}{!}{\includegraphics{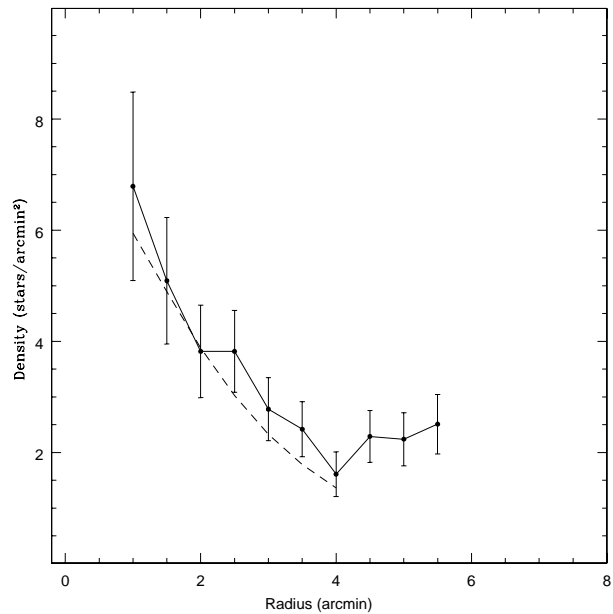}}
\caption{The radial density of the cluster as a function of the
radius is shown in bold line. The dotted line shows the
fitted profile.}
\label{figure3}
\end{figure}

The RDP is fitted with the function
$ \rho(R) \propto f_0/(1+(R/R_0)^2)$, where $R_0$ is the core radius, at which the
density $\rho (R)$, becomes half of the central density, $f_0$.
It was not able to
fit the RDP estimated with stars brighter than V = 19 mag, which indicates
its uneven nature. The best fit to the profile as shown in the figure 3. 
Considering that this is the best possible estimate, the core radius
of the cluster was estimated as 2.4 arcmin. The fitted profile as shown in the figure
also indicated that the density profile beyond the radius of 4 arcmin, shows a significant
rise. The radius at which this is observed is shown as big circle in figure 1. A clear 
decreament in the density of bright stars towards the cluster radius can be observed 
along with the presence of
bright stars just beyond the circle. This may indicate a healthy population
of bright stars in the field of the cluster, likely to be caused by spatially extended star formation.

\begin{figure}
\resizebox{\hsize}{!}{\includegraphics{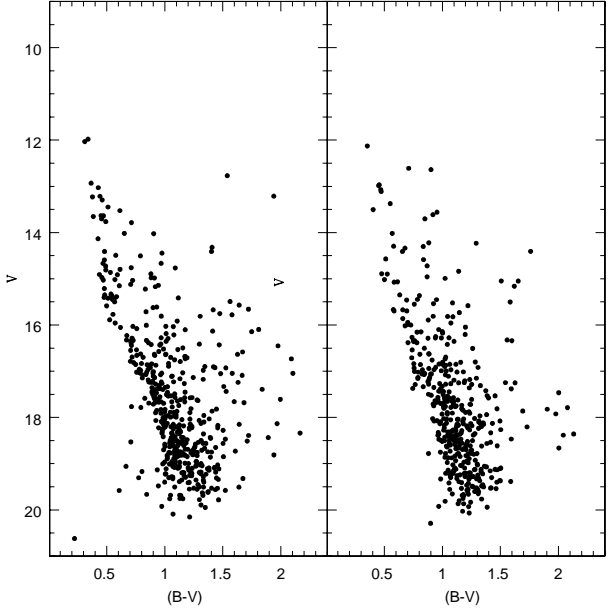}}
\caption{The V vs (B$-$V) CMDs of the stars located in the cluster region (left plot) and the field region
(right plot) are shown.
}
\label{figure4}
\end{figure}
\begin{figure}
\resizebox{\hsize}{!}{\includegraphics{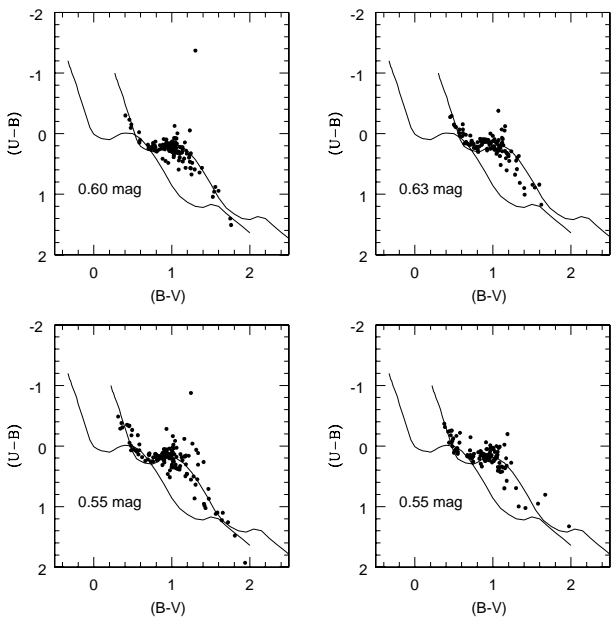}}
\caption{The ZAMS fit to the (U$-$B) vs (B$-$V) diagrams for the observed regions are shown here.
The four diagrams represent the four quarters of the observed region. The
estimated reddening and distance modulus for each region are indicated.}
\label{figure5}
\end{figure}

\subsection{Estimation of differential and average reddening} 
The colour-magnitude diagram (CMD) of stars located inside the cluster and field
are shown in figure 4. The stars located inside the radius of the cluster are used to
make the cluster CMD, whereas stars located outside are used to make the field CMD.
The cluster CMD (left panel) shows a well defined main-sequence 
with stars populated up to V=12 mag. There is also a healthy population of
stars located to the right of the MS. These stars may be field stars or PMS stars
considering the young age of the cluster. The field CMD also shows the presence of a 
good number of bright stars. 

The amount of interstellar reddening towards this cluster is estimated
using the (U$-$B) against (B$-$V) colour-colour diagram (CCDMs). Since the cluster
is known to be very young, it is likely that the region of the cluster has variable
reddening.
The study by Phelps \& Janes (\cite{pj94}) indicated the presence of differential 
reddening across the face of the cluster.
In order to estimate the amount of differential reddening, we divided the
observed region into 4 equal parts. The CCDMs for stars within
each region were obtained and fitted with unreddened ZAMS (Schmidt-Kaler 1982).
The corresponding CCDMs are
shown in figure 5, fitted with ZAMS. The estimated values of 
E(B$-$V) are also shown.

It can be seen that there is a fair amount of differential
reddening across the face of the cluster with the minimum 
and the maximum values of reddening being 0.63 and 0.55 mag respectively. 
It must be noted that the values estimated 
for each region may also be an average in itself. Further refinement in the 
estimate is not possible as the number of stars per region becomes too small 
to reliably estimate reddening.  
The reddening towards the lower half of the
observed region is 0.55 mag and this is where most of the cluster stars are located,
as seen from the figure 1. The upper half is more reddened and is populated mostly
by field stars. Therefore,
the mean reddening towards the cluster is estimated
to be E(B$-$V)=0.55 $\pm$ 0.04 mag. The error in the estimation of reddening
for each region is 0.02 mag and the average error in the estimation of (B$-$V) colour is
0.03 mag. These errors are quadratically added to obtain the error quoted above.
The reddening towards the field region is estimated to be 0.62$\pm$ 0.04.
Though the reddening values are within the errors, the marginal reduction in the
reddening near the cluster region may be due to the material being driven away by the massive stars
in the cluster.

\section{Distance to the cluster}
\begin{figure}
\resizebox{\hsize}{!}{\includegraphics{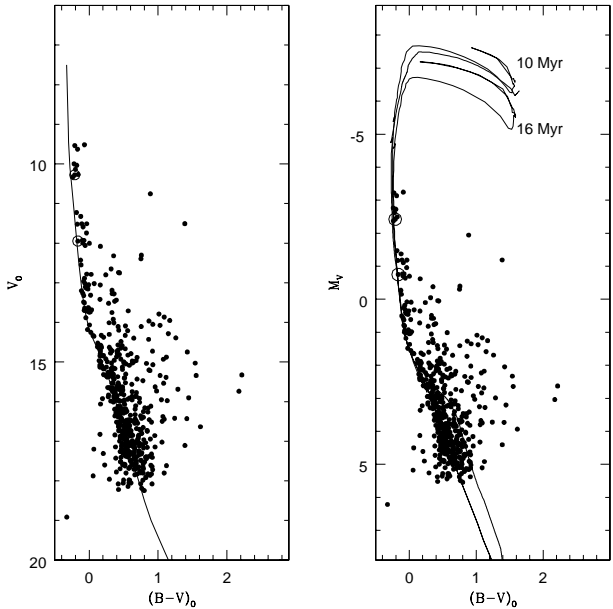}}
\caption{The ZAMS fit to the unreddened cluster MS in the V vs (B$-$V) CMD is shown in the 
left panel.  The extinction corrected magnitudes are plotted on the y-axis. The distance 
modulus is estimated to be 12.7 mag. The age of the cluster is estimated by fitting the 
isochrones to the turn-off of the MS as shown in the right panel. The age of the fitted 
isochrones are indicated. The location of two Be stars are shown. }
\label{figure6}
\end{figure}
The de-reddened cluster CMD is fitted with the ZAMS to
estimate the distance modulus, as shown in figure 6.
The absolute distance modulus is estimated to be 12.7$\pm0.2$ mag. 
This corresponds to a distance of 3470$^{+335}_{-305}$ pc, which can be
considered as 3500 pc. Phelps \& Janes (\cite{pj94}) obtained a reddening of 0.70 mag
and a distance of 4786 pc. Their reddening and distance estimates are much larger
than those estimated here. On the other hand, Kimeswenger \& Weinberger (1989)
estimated a distance of 3370 pc, which is in good agreement with the present estimates.

\section{Estimation of age of the cluster}
The estimation of age of the cluster can be done in many ways. First we use the MS turn-off to
estimate the age using isochrone fitting as well as synthetic CMDs.

The isochrones from the Padova group, Bertelli et al. (\cite{b94}) are used to estimate the age.
There are a few bright stars present in the cluster, for which we could not estimate the magnitudes,
as they were saturated. The magnitudes of these 7 stars are taken from the literature.
Magnitudes of 6 stars are taken from  Phelps \& Janes (\cite{pj94}) and one from Jasevicius (1964).
These stars are used to make the upper MS complete. The isochrone fit to the CMD is
 shown in figure 6.
The age of the cluster is estimated from the brightest stars on the MS, since there are no
evolved stars. Two isochrones of age 10 and 16 Myr are found to fit the upper MS. Thus we
estimate the younger age limit of the cluster as 10 Myr, whereas 16 Myr is also equally probable.
The cluster cannot be younger than 10 Myr, whereas it could be slightly older. The older limit
could not be estimated as there are no red giants or super giants.
The presence of Be stars in the upper MS indicate that the age of the cluster is similar to 
the above value, as Be stars are generally
seen in cluster in the age range of 12 -- 25 Myr (Meader et al. 1999). Phelps \& Janes (\cite{pj94}) found that
the cluster is younger than 10 Myr. 

\begin{figure}
\resizebox{\hsize}{!}{\includegraphics{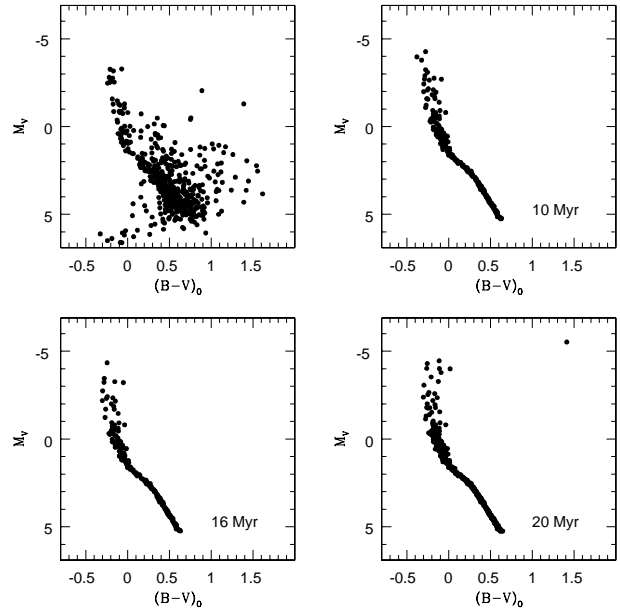}}
\caption{The observed and the synthetic CMDs for ages 10, 16 and 20 Myr are shown.
}
\label{figure7}
\end{figure}
We also estimated the age of the cluster by comparing the features in the observed CMD with synthetic CMDs
of different age.
It is seen that the cluster CMD is better understood when compared with a simulated or synthetic
CMD made from the evolutionary models than with isochrones. A synthetic CMD
in general distributes stars along the isochrone corresponding to the age of the cluster,
in accordance with the
time-scale of each evolutionary state in the CMD.
We constructed synthetic CMDs using the theoretical 
stellar evolutionary models presented by Bressan et al. (\cite{b93}). We used
solar metallicity models to create CMDs. The algorithm used to make 
synthetic CMDs are
presented in Subramaniam \& Sagar (1995) and also used in Subramaniam \& Sagar (1999).
We included the effects due to binaries with mass ratio between 0.75 -- 1.25 and also
the photometric errors in V and (B$-$V). Salpeter value for the mass function,
which is 2.35 (Salpeter 1955), is assumed in general. 
A control parameter is required to create synthetic CMD, so that it can be
directly compared with the observed CMD. For this, one can either use the number of red giants, 
or the number of stars near the tip of the MS, below the turn-off. Since there are no red giants
in this cluster, we used the stars near the top of the MS for this purpose.
In the observed CMD, there are 8 stars between 9.5 -- 10.5 in V magnitude.
This is used as the control number. Initially we used a value of 30\% for the fraction
of binary stars
in the cluster, the resulting CMDs were seen to have too many stars near the evolved
part of the CMD, close to the binary isochrone path. The value of
binary fraction is assumed to be zero as even a small value was found to widen the upper MS.
Thus the cluster has a very low fraction of binaries, at least among the more massive
stars populating the upper MS.
Figure 7 shows the observed and the synthetic CMDs for 3 ages, 10, 16 and 20 Myr.
In the synthetic CMD of age 10 Myr, there are stars brighter than the tip of the 
MS in the observed CMD. In the 16 Myr CMD, the stars in the MS look
more or less similar. For ages more than 16 Myr, the synthetic CMDs are found to have red giants and their
number was found to increase with age. Therefore, the method of synthetic CMDs also constrain the
age of the cluster in the range 10 -- 16 Myr.

\section{Pre-MS stars and turn-on age}
\begin{figure}
\resizebox{\hsize}{!}{\includegraphics{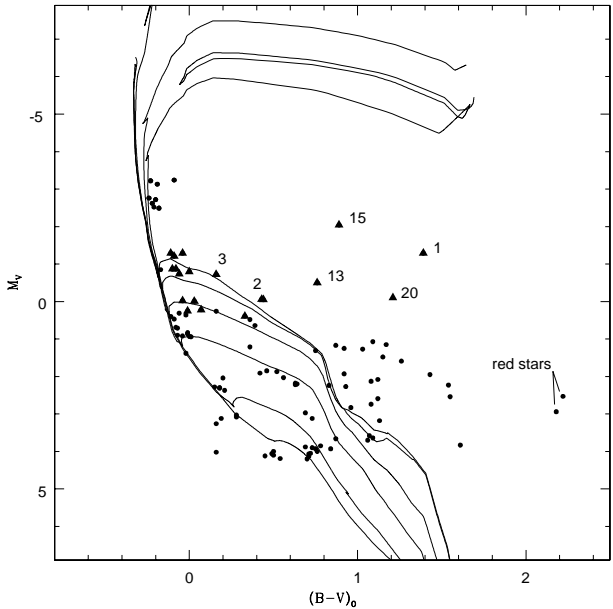}}
\caption{The age of the cluster as estimated from the MS turn-on and the
PMS stars, after correcting for the field stars. The PMS isochrones from Palla \& Stahler (1993)
for 0.1, 0.3, 1.0, 3.2, 10, 18 Myr are shown. The triangles show the stars for which spectra are obtained.
}
\label{figure8}
\end{figure}
Since the cluster is found to be young, it is possible that many of the low mass stars
are still in the PMS phase. These stars in the CMD occupy the same area as the field stars.
Hence, it is not easy to identify a PMS star in a CMD. If PMS stars are present in
the cluster, then these stars can be used to estimate the turn-on age of the cluster using
PMS isochrones. Therefore, it is important to identify the PMS stars, if they are present.
This can be achieved if we somehow identify and remove the field stars present in the region of the cluster.
The region outside the circle as shown in figure 1 is used for this purpose.
This region is considered as the control region to remove the field stars. This region is
located very close, just outside the cluster radius. Therefore, it is quite possible that some
cluster members, especially the low mass stars are present in the assumed field region. This 
would result in an over subtraction
of low mass stars in the CMD. Since our aim is to identify the presence of PMS stars and not  to estimate the
statistics of a complete sample, the choice of the control region is justified.

The field stars from the
cluster CMD are removed using a technique called zapping technique. 
This procedure was
used in the analysis of LMC clusters by Subramaniam \& Sagar (1995).
As the procedure adopted for field star removal
is a statistical one, the stars which have the maximum probability to be a field star
in the cluster CMD are removed. Therefore the probability of the presence of field stars
in the final CMD is very less, though a few field stars still may be present.
The field subtracted CMD is presented in figure 8 along with the PMS isochrones. 
The PMS isochrones are taken from Palla \& Stahler (1993).
The cluster could have a very young turn-on age of about 0.1 Myr as seen from the figure.
There is hardly any star located on the MS below the turn-on point corresponding to the 3.2 Myr isochrone. 
Therefore, it is possible that the turn-on age of this cluster is about 3 Myr.
In order to identify the other known PMS features in the above mentioned stars, we 
used low resolution spectra of the candidates obtained using the slit-less mode of
observations and available JHK data as described in the following sections.
\section{Spectra of Be stars and some probable PMS stars}
\begin{figure}
\resizebox{\hsize}{!}{\includegraphics{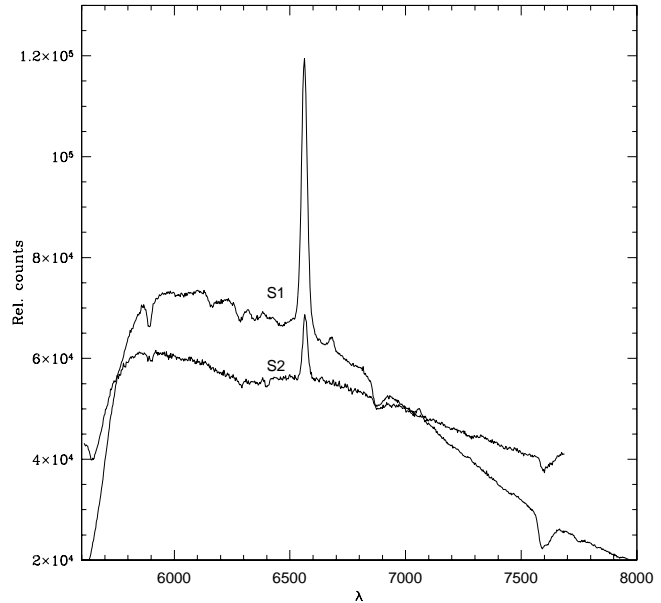}}
\caption{Spectra of two stars identified as Be stars are shown.
}
\label{figure9}
\end{figure}
\begin{figure}
\resizebox{\hsize}{!}{\includegraphics{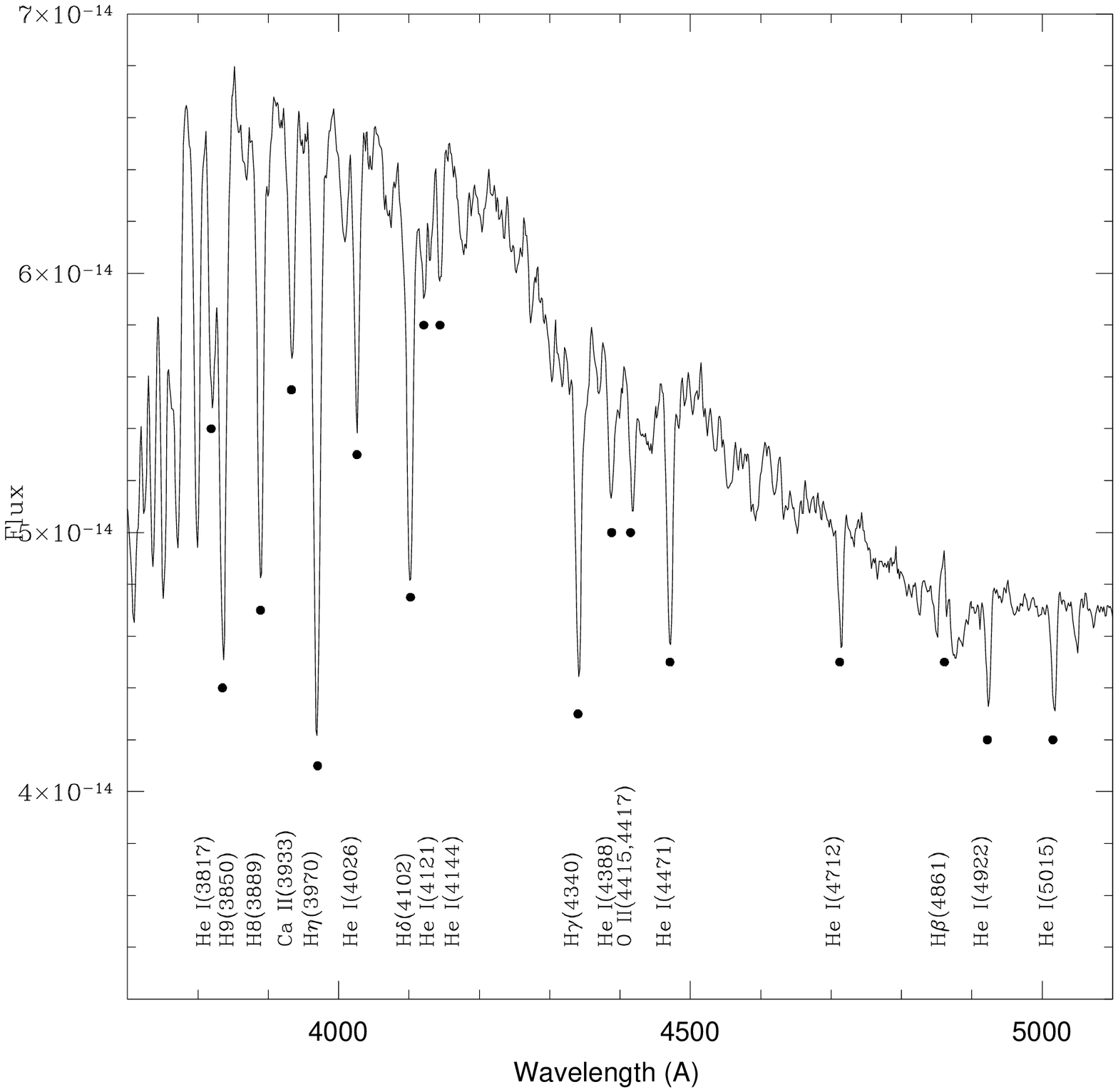}}
\caption{The blue spectrum of the bright emission star S1.
}
\label{figure10}
\end{figure}
\begin{figure}
\resizebox{\hsize}{!}{\includegraphics{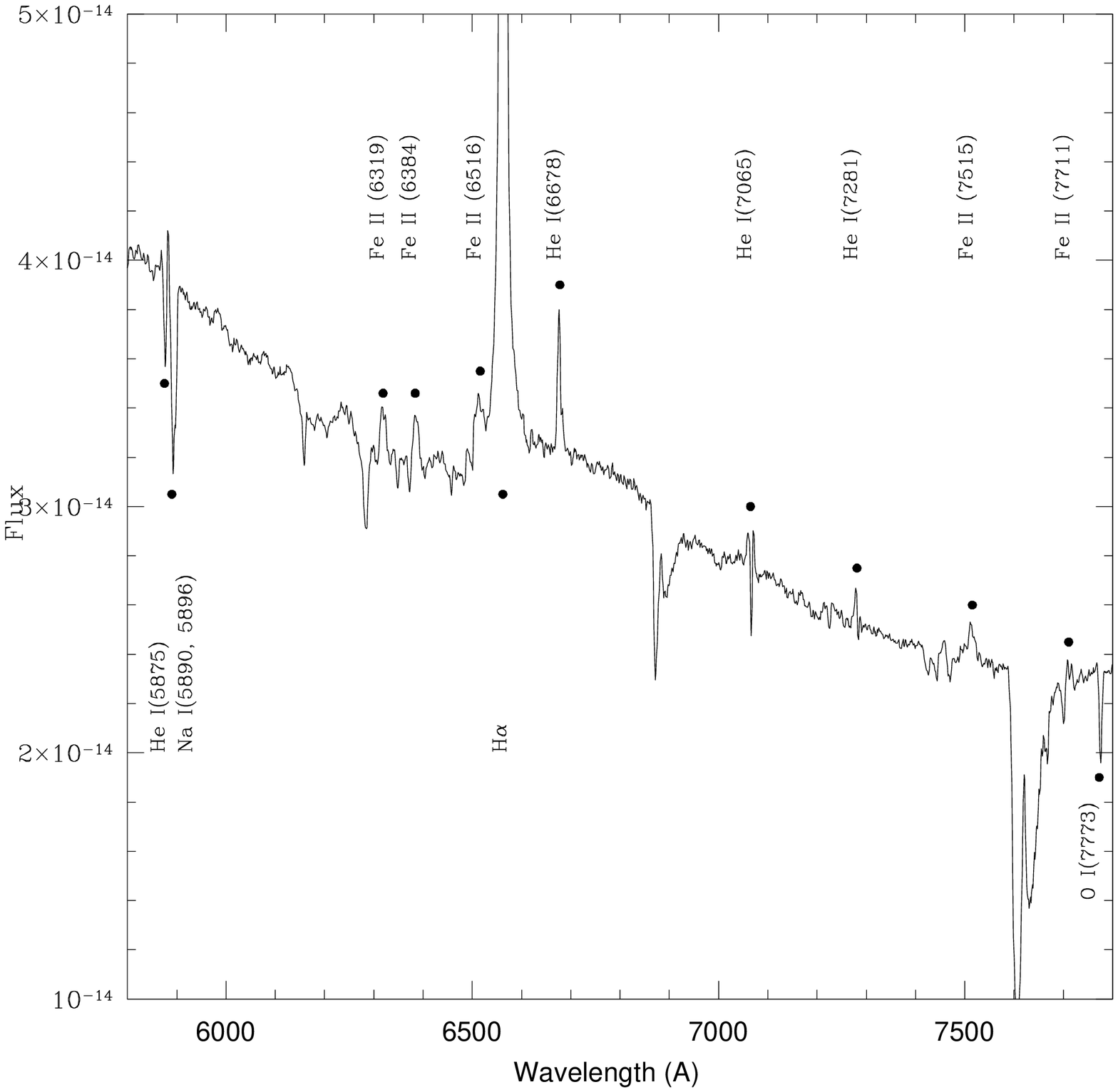}}
\caption{The spectrum of the bright emission star S1.
}
\label{figure11}
\end{figure}
\begin{figure}
\resizebox{\hsize}{!}{\includegraphics{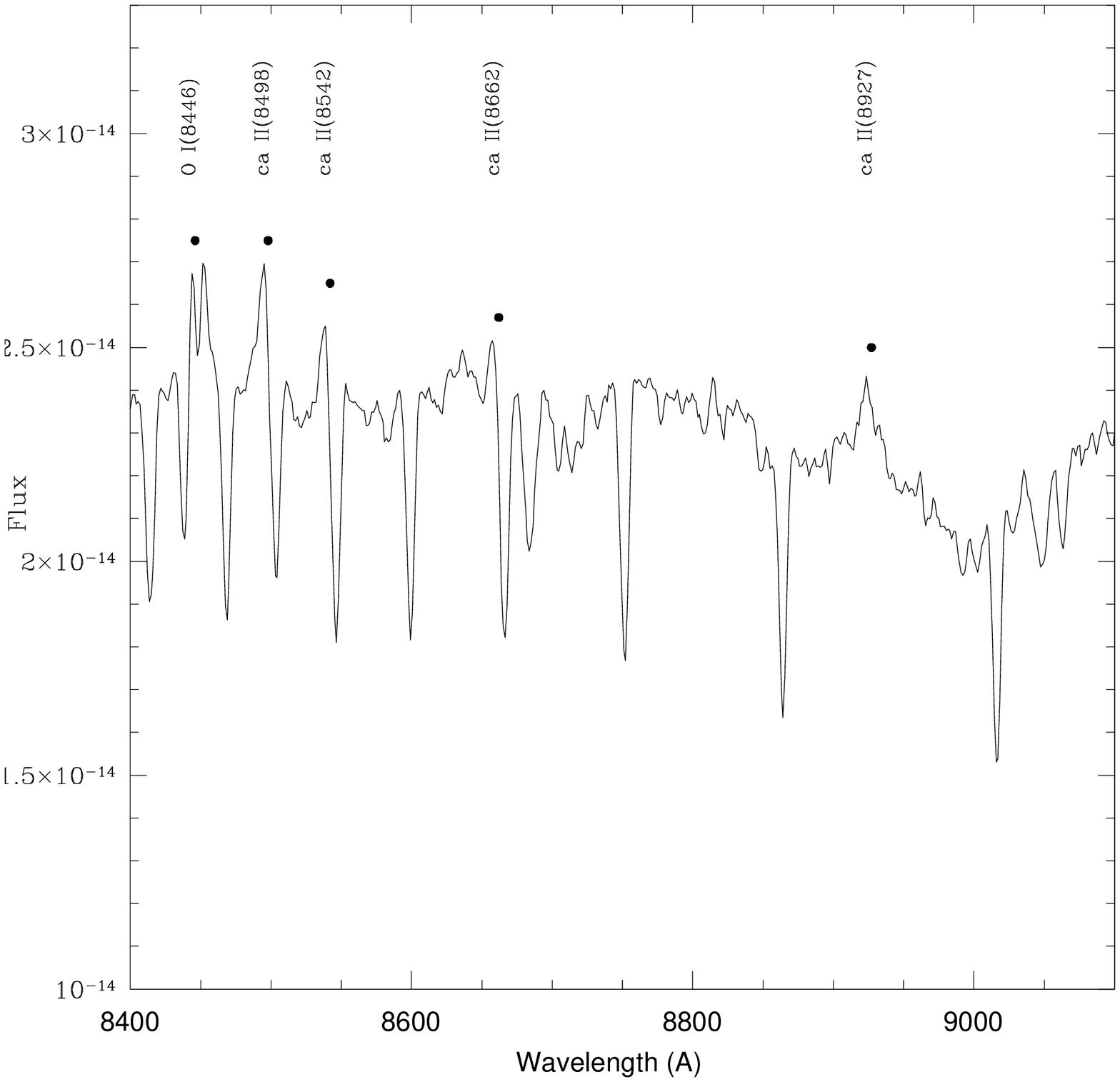}}
\caption{The red spectrum of the bright emission star S1.
}
\label{figure12}
\end{figure}
\begin{figure}
\resizebox{\hsize}{!}{\includegraphics{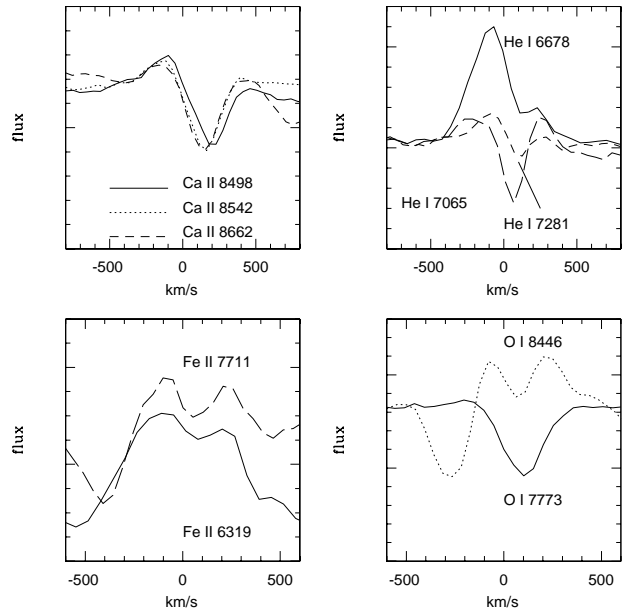}}
\caption{The line profiles of star S1.
}
\label{figure13}
\end{figure}

\begin{figure}
\resizebox{\hsize}{!}{\includegraphics{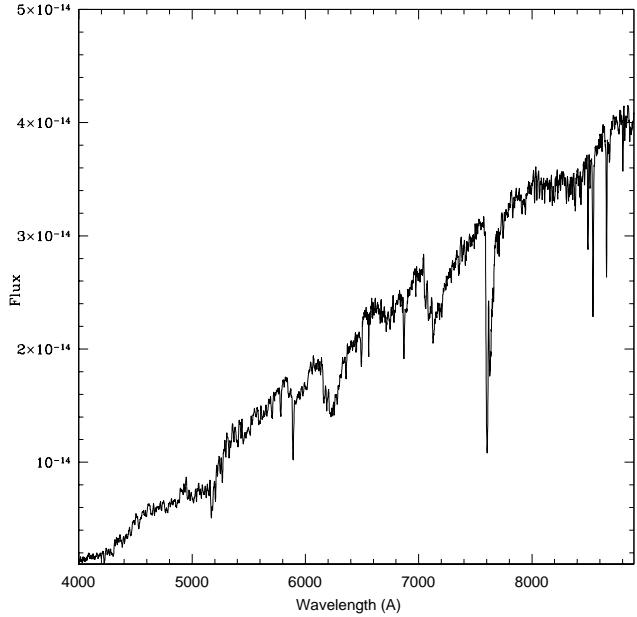}}
\caption{Spectrum of star 1.
}
\label{figure14}
\end{figure}
\begin{figure}
\resizebox{\hsize}{!}{\includegraphics{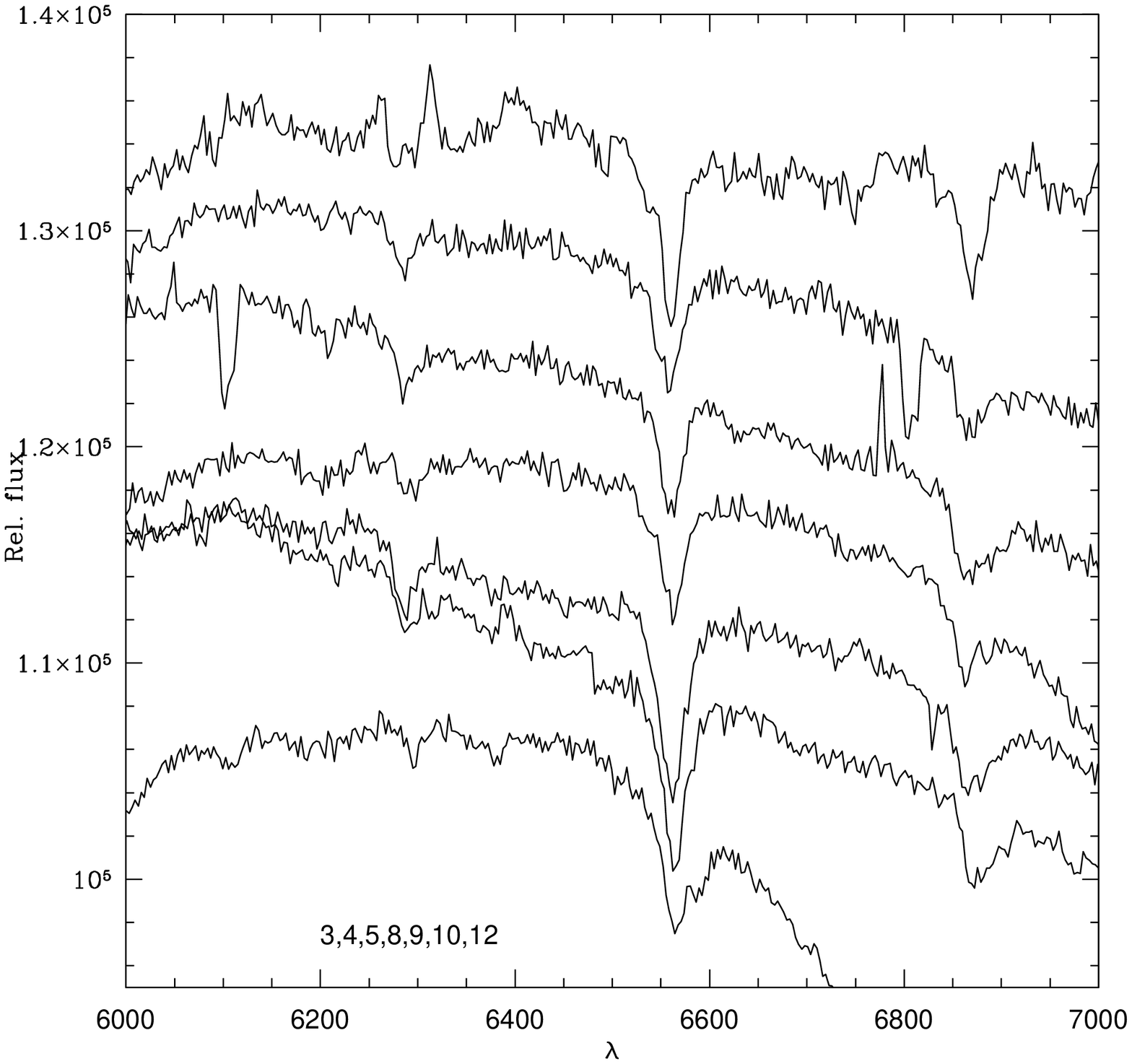}}
\caption{Spectra of 7 probable PMS stars are shown.
}
\label{figure15}
\end{figure}
\begin{figure}
\resizebox{\hsize}{!}{\includegraphics{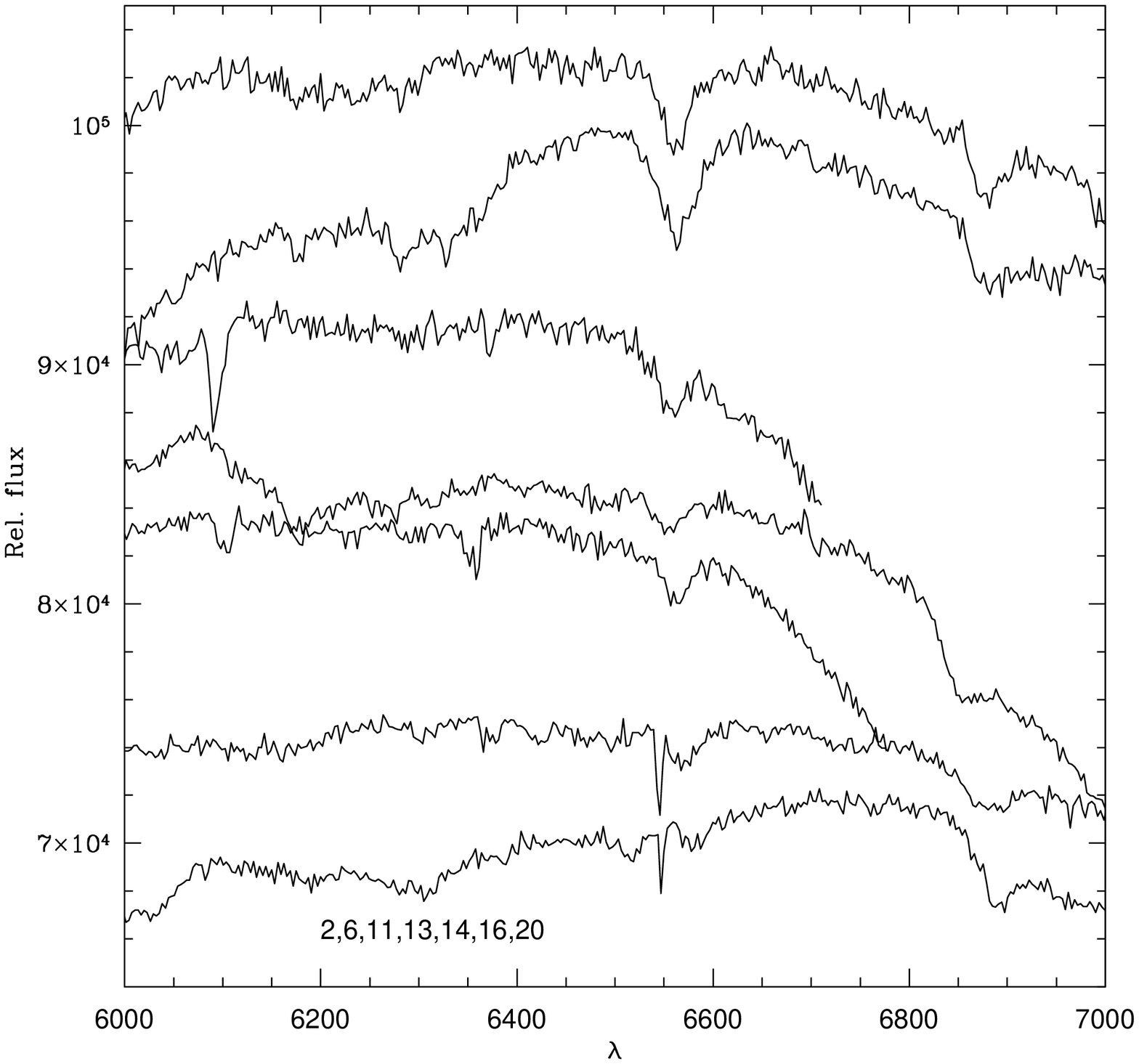}}
\caption{Spectra of 7 probable PMS stars are shown.
}
\label{figure16}
\end{figure}
\begin{figure}
\resizebox{\hsize}{!}{\includegraphics{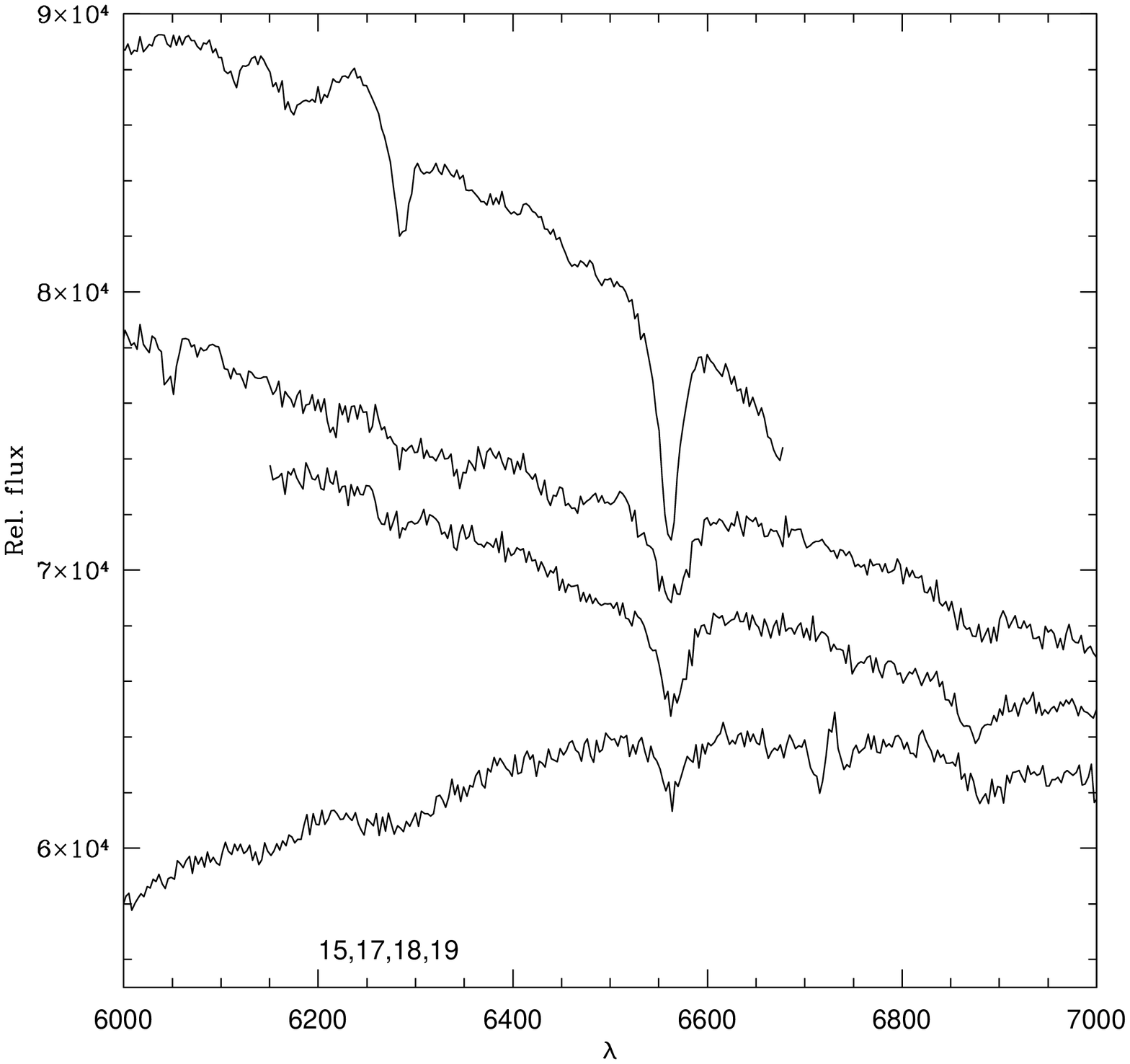}}
\caption{Spectra of 4 probable PMS stars are shown.
}
\label{figure17}
\end{figure}

\begin{figure}
\resizebox{\hsize}{!}{\includegraphics{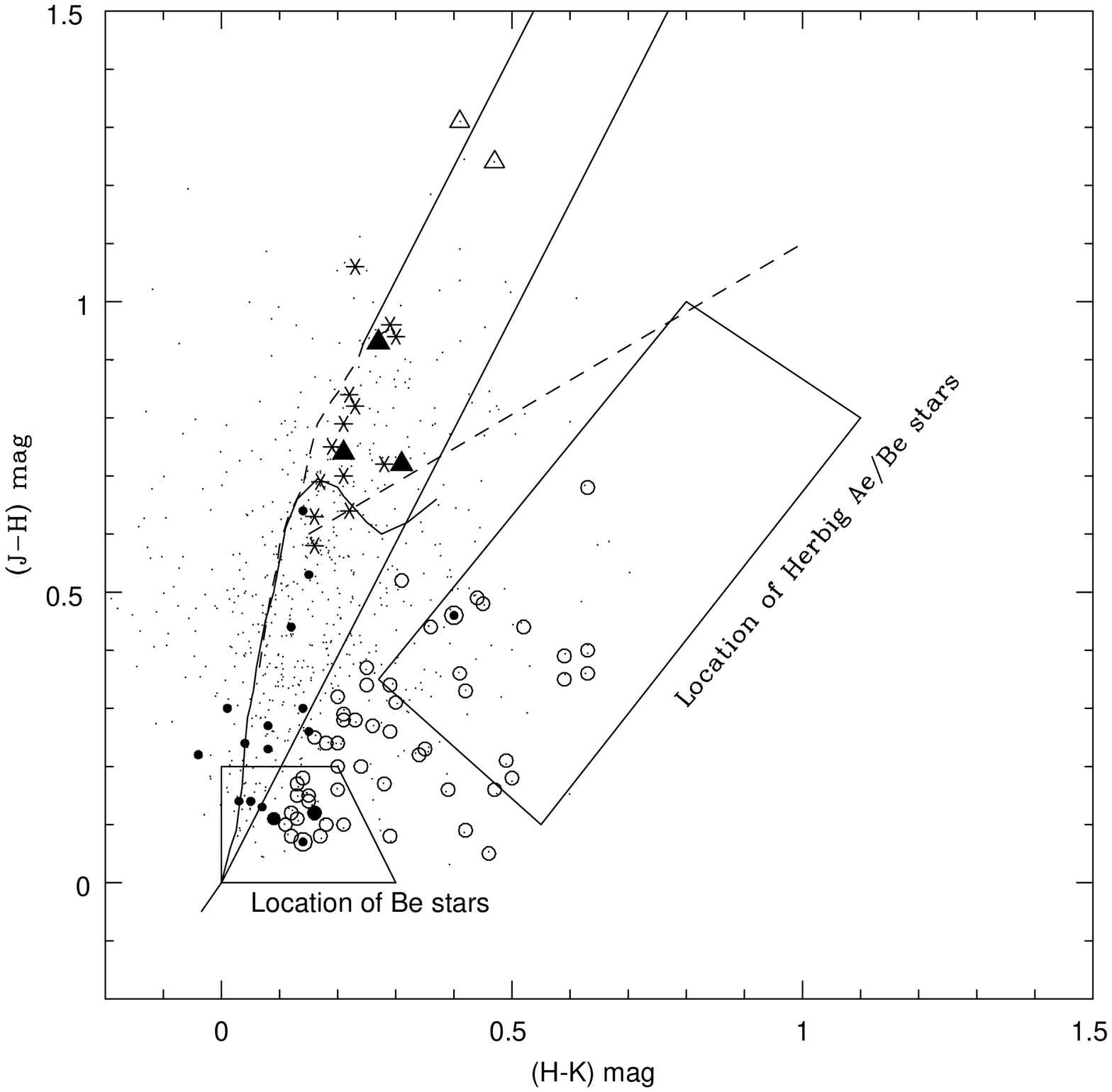}}
\caption{(J$-$H) vs (H$-$K) colour-colour diagram of stars in the field of NGC 146 is shown.
The filled circles denote stars for which spectra were obtained. The two H$\alpha$ emission
stars are denoted by dot with circle around them. The filled triangles are the PMS stars
with high extinction. The symbol star denotes objects which are located to the right of
the PMS isochrones. The MS and the giant star locations in the CMD are shown 
(Bessell \& Brett 1988). The location of T-Tauri stars are shown as the dashed straight 
line (Meyer et al. 1997). The location of Be stars is taken from Dougherty et al. (\cite{dou94}) 
and the location Herbig Ae/Be stars is taken from Hernandez et al. (\cite{h05}).
}
\label{figure18}
\end{figure}
We obtained slit-less spectra of stars in the cluster with the aim of
identifying emission line stars. This resulted in the identification of
two Be stars, whose spectra are shown in figure 9. 
The brighter of the two, S1 shows a few more lines in emission other
than H$_\alpha$. The helium lines at  6678 \AA and 7065 \AA are in emission.
Some of the Fe II lines may also be in emission. The above features,
indicate that hot circumstellar material could be present around S1. The spectral
type of S1 is found to be earlier than B3. The other Be star, S2 is fainter and its
spectral type is found to be later than B5. In order to study the spectral features of
S1 in detail, we obtained slit spectra of S1 between 3500 -- 9200 \AA, on 18 January 2005,
to confirm the He and Fe emission. The flux calibrated spectrum of S1 is shown in 
figures 10, 11 and 12. It can be seen that in the blue region, a number of He I lines 
are seen in absorption. The H$\beta$
line is found to show a P-Cygni profile. In figure 11, we identified 5 Fe II lines and
3 He I lines in emission, apart from the H$\alpha$ emission. The Fe II and He I lines are
also found to show structures. In the red spectrum, shown in figure 12, it can be seen that the
O I ( 8446\AA) is in emission whereas the O I (7773\AA, figure 11) is in absorption.
Another interesting feature is the inverse P-Cygni profiles of the three Ca II lines 
in figure 12.
This feature is found to be similar to KK Oph and XY Per (Hamann \& Presson \cite{hp92}), which are
Herbig Ae/Be stars.
This is a signature of gas infall. The line profiles are shown in figure 13,
and were used to estimate the radial velocity of the star and the infall velocity.
The structures seen in He I and Fe II may also
be due to the self absorption in the line of sight as a result of the infalling gas.
Just as in the case of KK Oph, the velocities were found to be lower in Fe II and 
higher in Ca II.
The radial velocity as estimated from the three Ca II emission profiles is found to be $-$120$\pm$25 kms$^{-1}$.
The difference in velocity between the emission and the absorption is assumed to be the
infall velocity. This was found to be +300 kms$^{-1}$ for the three Ca II lines and +150 kms$^{-1}$ for 
two Fe II and one O I lines. 

The slit spectrum of the red star in the CMD denoted
as 1 is shown in figure 14. This spectrum was also obtained on 18 January 2005. 
The spectrum indicates that it is a cool star. The strong absorption bands and the absorption
feature at $\sim$ 6500 \AA could indicate that the spectral type is similar to an M 2 giant.
The absorption feature at $\sim$ 6500 \AA\ arises from a blend of Fe I, Ca I and 
Ba II lines and is also seen in FU Orionis type stars (Mundt et al. 1985). The spectrum 
looks very similar to that of PP 13N as shown in the figure 5, Aspin \& Sandell (\cite{as01}), 
except for the H$_\alpha$ emission. They classified PP 13N as an M 2.5 giant in the PMS phase. 
Therefore, it could be a PMS star or a field star.
We estimated the radial velocity using the three Ca II lines and was found to be 
$-$140$\pm$25 kms$^{-1}$. This value is within the errors of S1. Thus this star is 
likely to be a cluster member.
We searched for fainter stars with H$_\alpha$ in emission in the slit-less spectra,
but could not find any. We repeated the observations on three more nights,
25 June 2004, 21 July 2004 and 23 November 2004. None of the stars brighter than V = 15.0 mag
showed H$_\alpha$ emission. We obtained two 
15 min exposures on 25 June 2004 and 21 July 2004 and co-added the images
to increase the signal. The seeing was poorer on 25 June 2004, which resulted in lower
signal than on 21 July 2004. Therefore, we used the 21 July spectra for further
analysis. We extracted the spectra of 18 probable PMS stars which are shown as
triangles in figure 8. Spectra of stars brighter than V $\sim$ 15 mag had good signal, whereas
the spectra of fainter stars had very poor signal. Hence we could only extract spectra
of 18 stars brighter than the above limit and are shown in figures 15, 16 and 17.
The magnitudes and location of these stars as in figure 1 are given in table 2.
These stars did not show H$_\alpha$ emission in any of the three observations.
It can be seen that the spectra are more or less featureless, since the
lines are not clearly formed. The noticeable feature is the presence of
very broad H$_\alpha$ line in absorption. The spectra are grouped together and 
presented, such that
spectra with deeper H$_\alpha$ are shown in figure 15, broader and shallower lined 
spectra are shown in figures 16 and 17. The spectrum of star 15 is shown in figure 17, 
although it does not belong to the group. Spectra of 3 and 5 indicate the likely hood of
some emission features.
Heske \& Wendker (\cite{hw85}) found that separating stars based on the narrow
and wide Balmer lines helped to identify the PMS stars. They argue that
on the way to the MS, the emission lines would slowly disappear in the
spectra of PMS stars, with the Balmer lines first. These will switch over to
broad absorption lines and then later to become typical Balmer lines.
Thus we do not find any clear signature, but only a probable signature of PMS phase in these stars.
In the young open cluster NGC 6611, de Winter et al. (\cite{dew97}) classified stars located redward
of the MS as different groups. Following this classification, stars 2 and 3 fall in group II and
1, 13, 15 and 20 fall in group III. 
Group II is suggested to be an interesting group of
likely PMS stars and group III of evolved or field stars or stars with large extinction.

\begin{table*}
\caption{The location and magnitudes of two Be stars and 19 probable PMS stars for which spectra are shown.
The J, H and K magnitudes are taken from 2MASS catalogue. The type of star, presence of NIR excess, presence
of clustering around the star and
the number of probable Herbig Ae/Be stars located close to the star are indicated.
}
\begin{tabular}{ccccccccl}
\hline
Number & X & Y & V  & (B$-$V) & J & H & K & Comments \\
\hline
 S1 & 550.33 &682.40 &11.98 &0.34 & 10.52 & 10.06 & 9.66& Probable Herbig Be star, clustering, 2 stars \\
 S2 & 1316.10& 616.00& 13.65&0.39 & 12.72 & 12.65 & 12.51&Be star , 1 star\\
  1 &534.26&165.35 &13.21  &1.94 & 9.36 & 8.43 & 8.16 & probable PMS star, large A$_v$ clustering\\
  2 & 918.10& 36.42& 14.45 & 0.98 & 12.16 & 11.63 & 11.48 & group II star, clustering \\
  3 & 953.52&204.68& 13.79 & 0.71  & 12.27 & 12.01 & 11.86 &group II star\\
  4 & 749.51&858.96& 13.76 & 0.49  & 12.63 & 12.50 & 12.43 & NIR excess\\
  5 & 866.91&958.25& 13.64 & 0.45  & 12.61 & 12.47 & 12.44 & \\
  6 &1090.01&930.72& 14.49 & 0.58  & 13.12 & 12.88 & 12.84 &\\
  8 &1387.88&991.64& 13.64 & 0.47  & 12.55 & 12.41 & 12.36 &\\
  9 &1304.90&851.16& 13.30 & 0.46  & 12.21 & 12.10 & 12.01 & NIR excess, clustering, 2 stars\\
  10 &1247.02&653.53& 13.22 & 0.44  & 12.24 & 12.10 & 12.05 &\\
  11 &1289.43&1100.30 &13.53&  0.61  & 12.11 & 11.88 & 11.80 &\\
  12 &1033.60&1130.94 &14.02&  0.65  & 12.44 & 12.14 & 12.13 & mild clustering, 1 star\\
  13 &1113.08&1422.52 &14.32&  1.41  & 11.72 & 11.08 & 10.94 &\\
  14 &1438.33&1096.30 &14.80&  0.62  & 13.32 & 13.10 & 13.14 &\\
  15 &1283.24&1020.19 &12.77&  1.54  & 9.66 & 8.92 & 8.71 & large A$_v$\\
  16 &1590.90&1162.93 &14.77&  1.09  & 12.64 & 12.20 & 12.08 &\\
  17 & 577.48&1150.43 &14.91&  0.59  & 13.49 & 13.37 & 13.21 &IR excess, mild clustering \\
  18 &1525.15&1081.88 &15.04&  0.72  & 13.33 & 13.06 & 12.98 &\\
  19 & 598.31&227.53 &14.90 & 0.88  & 13.1 & 12.8 & 12.66 &\\
  20 & 336.86& 1013.99& 14.41& 1.76 & 11.04 & 10.32 & 10.01 & clustering, large A$_v$ \\
\hline
\end{tabular}
\end{table*}
\section{Near Infrared colour-colour diagram}
The near infrared (NIR) J, H, K magnitudes of stars in the region of the
cluster is available from the 2MASS catalogue. The K vs (J-K) CMD for 
stars located within 5 acrmin radius shows that a number of stars 
have redder colours, probably due to increased extinction.
The (J$-$H) vs (H$-$K) colour colour diagram is
shown in figure 18. The prominent feature seen is the presence of a large
number of stars showing NIR excess. These stars are located below the T-Tauri 
location indicating that these stars are more massive and probably belong to the 
class of intermediate mass PMS stars.
In order to group these stars, we have shown the typical location of Be stars 
(Dougherty et al. \cite{dou94}) and Herbig Ae/Be stars (Hernandez et al. \cite{h05}).

The H$\alpha$ emission star S1 is found to be located in the Herbig Ae/Be location 
supporting the evidence obtained from the spectra that it is most likely to be a Herbig 
Be star. On the other hand, the star S2, is located in the Be region. The stars which were 
not found to show H$\alpha$ emission in 
their spectra,  4, 5, 9, 10 and 14 were found to be located in the Be region and 4, 9 and 17
have NIR excess. 18 stars are found to have NIR excess and located in the region of Be stars
in the colour-colour diagram. The location of Herbig Ae/Be stars is also well populated. 
We found 17 stars, including S1, to be located in this region. This may be one of the very 
few clusters hosting such a large number of intermediate mass PMS stars. We also notice 
the presence of a large number of stars located between the Be and the
Herbig Ae/Be location. It will be very interesting to find out the nature of these stars.
On the whole, inside the cluster radius, we have found 54 stars showing NIR excess.
Three stars, 1, 15 and 20 (filled triangles) are found to be located
in a region indicating large extinction. In the field subtracted CMD, we found the presence 
of stars which are located to the right of the PMS isochrones. These stars (star symbol) 
are found to have
very large extinction. The two red stars in the optical CMD are shown as open triangles, which show
the largest extinction. Thus a good number of stars are found to have large extinction.
We would like to point out that no optical counterpart could be found for a number of stars
identified in the NIR. This is a clear indication of the presence of embedded stars.
Thus, NGC 146 shows the presence of a large number of probable intermediate mass PMS stars,
embedded stars and stars with large extinction. 

\begin{figure}
\resizebox{\hsize}{!}{\includegraphics{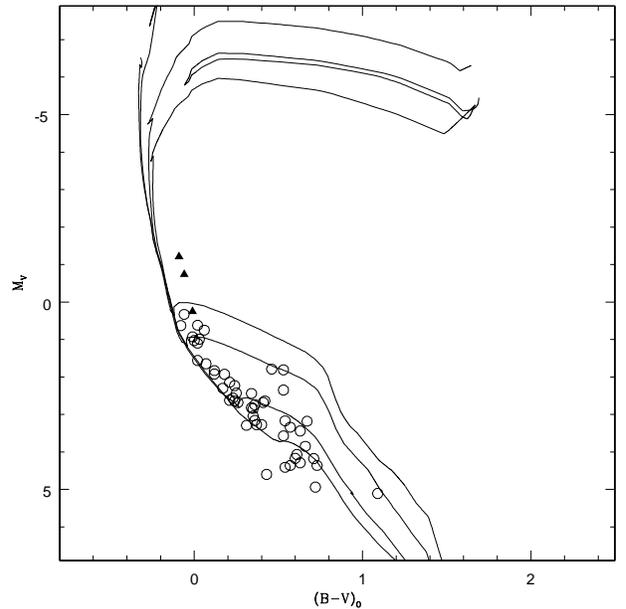}}
\caption{The optical CMD with stars with NIR excess in the region of the cluster. The 
PMS isochrones are shown
for ages 1, 3.2, 10 and 18 Myr. The symbols have the same meaning as explained in figure 18.
}
\label{figure19}
\end{figure}
The optical M$_V$ vs (B$-$V)$_0$ CMD of the cluster is shown in figure 21, where the stars
with NIR excess within the cluster region is shown. The stars are referred to by the same 
symbols as in the earlier diagrams. {\bf Some of these stars are not present in figure 8, since
they could have been identified as field stars. In this figure, we try to identify the
population of PMS stars with NIR excess and located inside the cluster radius. Therefore,
we have not attempted any field star removal.}
It can be seen that stars with NIR excess are more or less absent above the 1 Myr isochrone.
Most of these stars (open circles) belong to spectral types later than A0. 
The number is found to increase near the turn on point of the 3.2 Myr isochrone.
Therefore, it is very likely that the turn-on age is $\sim$ 3 Myr. This is similar to the value
obtained from the field star subtracted optical CMD (figure 8). It is also clear that  a good fraction
of stars located in the cluster region is very young.

\section{Clustering around the Herbig Ae/Be star}
\begin{figure*}
\centering
\includegraphics[width=17cm]{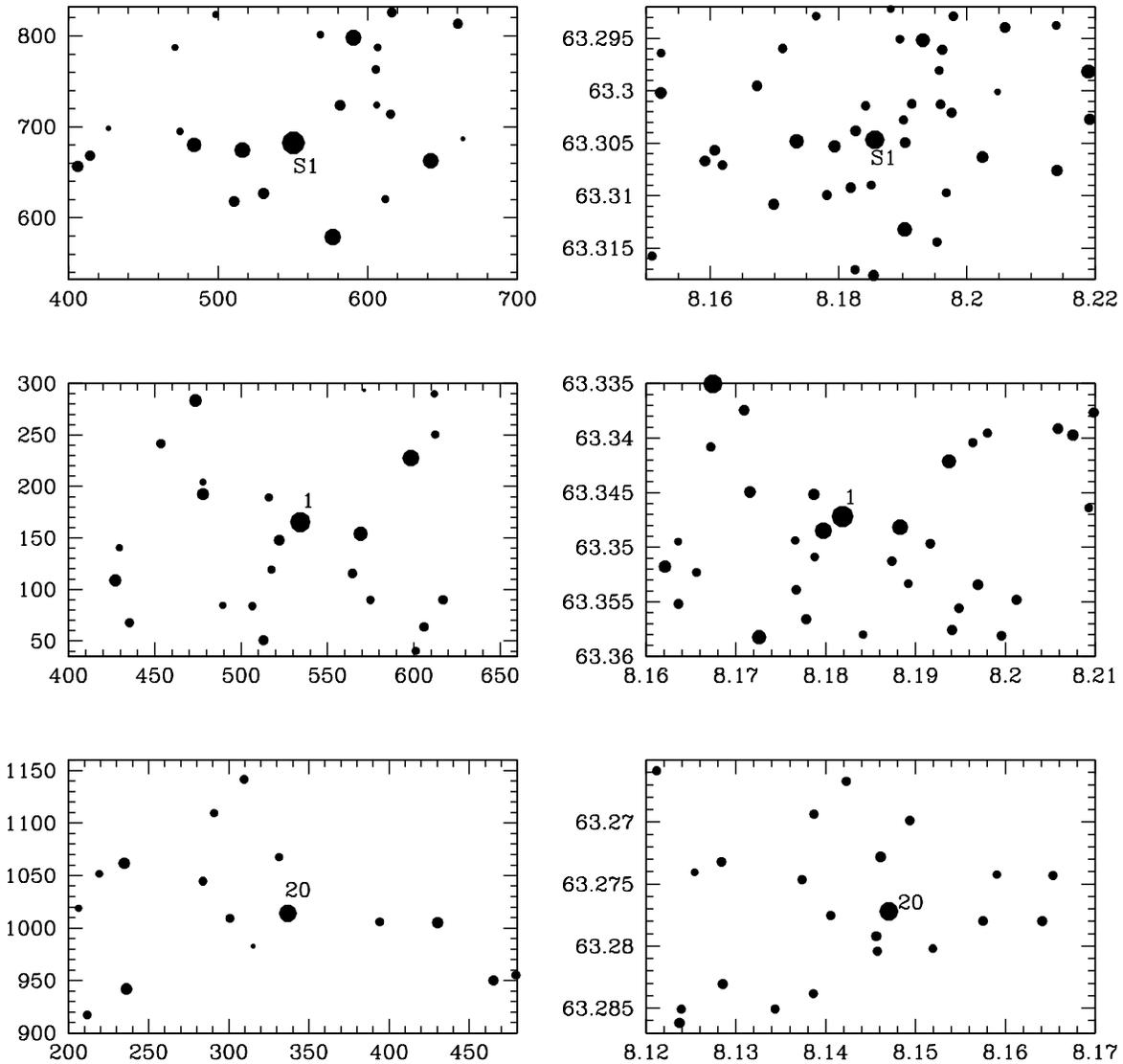}
\caption{The V passband (left panels) and the K passband (right panels) are shown around
the stars S1, 1 and 20. The K band data is from 2MASS.{\bf The axes are in 
pixels (1 pixel = 0.297 arcsec)
for the V band and in RA (J2000) \& Dec (J2000) for the K band.} A local 
enhancement in the stellar density can be noticed in K band images around these stars.
}
\label{figure20}
\end{figure*}
Multi wavelength studies of the environments of several Herbig Ae/Be stars in the optical,
NIR and millimeter (Barsony et al. \cite{bsk91}, Aspin \& Barsony \cite{ab94},
Palla et al. 1995) have shown that young clusters are still partially
embedded in the parent molecular clouds and that NIR observations especially in K-band are
best suited to detect the less massive companions to the Herbig Ae/Be star itself.
Hillenbrand et al. (\cite{h92}) found evidence for a correlation between the mass of the Herbig Ae/Be star
and the surface density of K-band stars detected around it. Testi et al. (1997) confirmed the
above results. Testi et al. (1999) found that early Herbig Be stars are surrounded by dense clusters
of low mass companions. In the present study, where we have found a new Herbig Be star, it will be
useful to know whether there is any local stellar density enhancement in its vicinity. We looked at the
immediate surrounding of the stars S1 and also some of the bright candidate PMS stars for stellar
density enhancement around them. We compared the optical V band data with the 2MASS K band data
and are shown in figure 20. The stellar density enhancement is found near the stars S1, 1 and 20.
Mild enhancements were found near the stars 12 and 17. 
The star S1 is surrounded by 5 stars which are visible in K-band and not in V-band. 
In the case of star 1, the stellar density is enhanced both in optical and K-band.
Around the star 20, four stars which are present in K-band do not have any counterpart in V-band.
In this cluster, it is interesting to note that
the densest stellar region is found between the stars S1 and S2. The stellar density to
the left of S1 and to the right of S2 is less than that between them.
Some of the probable PMS stars of lower mass were
located close to the higher mass stars. These are indicated in table 2. That is, two stars
each were found to be located
close to S1 and 9 whereas one star each is found to be located close to S1 and 12.

\section{Discussion}
The cluster is found to have an older turn-off (10 -- 16 Myr) and a younger turn-on age ($\sim$ 3 Myr).
The majority of the stars within the cluster radius is likely
to belong to the younger population as indicated by the large number of candidate
intermediate age PMS stars (54 stars). This region might have experienced a earlier epoch of
star formation between 10 --16 Myr and another recent one, when majority of the stars
are formed. A similar case found in the cluster NGC 6611 (Sagar \& Joshi 1979, Hillenbrand et al. \cite{h93}).
In this cluster, at least one star is found to be as old as 6 Myr and the age of the 
PMS population is $\le$ 1 Myr.
In the case of NGC 4755, the massive stars in the cluster are found to be at least 4 Myr older than
the bulk of the low mass stars (Sagar \&  Cannon 1995). It is possible that the star formation 
has continued in these clusters for at least 6--7 Myr.
It is also likely that multiple epochs of star formation is present in NGC 146.

The low resolution spectra have helped in identifying a few high mass (B type)
stars which do not show any emission feature, but have infrared excess similar
to the Be stars.  These stars were characterised by broad H$_\alpha$ absorption lines
are located in the region of Be stars in the colour-colour diagram. Manoj et al. (2002) found
three similar stars located in the Orion nebula cluster, which have similar NIR excess but do not
show any emission lines in their spectra. This might be due to the fact that the early B-type 
stars get rid of their disks and most of the circumstellar material very fast so that the 
emission features are short lived (Hillenbrand et al. \cite{h93}). Delgado et al. (\cite{del99})
did not find any H$_\alpha$ emission in their candidate PMS stars  in IC 4996, which is found to be
$\sim$ 8 Myr old. In the case of NGC 146, 
there is only one star, S1 which is not only found to retain material around it
but also found to show signatures of accretion as well as clustering around it.
 
It will be interesting to find out the extent of star formation near this cluster.
A study of a larger region around this cluster in optical and NIR is ideal to
bring out the star formation history in the Perseus arm around this cluster.
Another young cluster, King 14, is found to be located close to NGC 146 and these are one of the candidates
for double open cluster pairs of Subramaniam et al. (1995). The slit-less spectral observations
of King 14 revealed the presence on 1 star showing H$\alpha$ in emission, which confirms its young
nature. If they are located at the same distance the lateral separation between these clusters
is less than 9 pc. It is important to study the region containing both these clusters in order to
find out if these clusters are  binaries.
There is no other star cluster located very close to NGC 146, though young clusters
like Markarian 50 (Baume et al. \cite{bvc04}) and NGC 7510 (Sagar \& Griffiths 1998; Barbon \& Hassan \cite{bh96})
are located in the Perseus arm and are similarly old. These clusters are located at $l \sim 110^o$ and 
similarly distant and hence
the lateral separation between NGC 146 and the two clusters is about 500 pc. We have observed
the above two clusters in the slit-less mode to search for emission line stars and we have found that
NGC 7510 hosts 2 and Markarian 50 hosts one star showing H$\alpha$ in emission. The clusters NGC 663 
and NGC 7419, which are
known to host large number of Be stars are also located in this direction, but about
1 kpc closer. Similarly the only known binary cluster pair NGC 869 and NGC 884, which also hosts
a number of Be stars is located in the Perseus arm.
Now NGC 146 is found to host a large number of probable Be as well as Herbig Ae/Be stars.
The question which naturally comes to the mind is that is there anything special about the region
which is forming clusters with a large number of Be stars. It will be interesting to look into this
aspect.
\section{Results}
The young cluster NGC 146 is studied using UBV photometry and low resolution spectra 
to estimate the cluster parameters. The cluster is found to have an older turn-off (10 -- 16 Myr) and
a younger turn-on age ($\sim$ 3 Myr). The difference between them being $\sim$ 7 Myr. 
A good fraction of the stars in the cluster region are found to belong to the younger age.
The cluster is found to be located at 3400 pc. Kimeswenger \& Weinberger (1989) presented
optical evidence of a spiral arm beyond the Perseus arm, in the second galactic quadrant.
NGC 146 was suggested to be part of this spiral arm, with a distance of 4800 pc (Phelps \& Janes \cite{pj94}).
The present distance estimate shows that the cluster
NGC 146 is located in the Perseus spiral arm itself as inferred from
figure 1a of Kimeswenger \& Weinberger (1989). Thus NGC 146 is a cluster located in the Perseus spiral arm
of the Galaxy.
{\bf The NIR colour-colour diagram using the 2MASS JHK photometry of stars in the cluster
region showed the presence of a large number of stars (54 stars) with NIR excess, stars with
large extinction and embedded stars indicating the youth of the cluster. This cluster is 
likely to have had either continued star formation for 7 Myr or multiple epochs of star formation.}

Two Be stars (one Be star and one Herbig Be star) are identified in the cluster for the first time. 
The Herbig Be star is found to be in the
B0-B3 spectral range and the other is found to be later than B5. We estimated the fraction of Be stars
in the B0 - B3 spectral range. There is one Be star out of 8 B stars in the above
range, after the removal of field stars. The fraction of Be stars is thus found to be 12.5\%.
This is similar to the fraction of Be stars found in NGC 581 (Maeder 1999). Also, NGC 581
is found to have an age of 16 Myr, very similar to the age of NGC 146.
In this study we found that the PMS stars in the
B spectral type did not show any H$_\alpha$ emission. They were characterised by
broad H$_\alpha$ absorption lines. Thus, the PMS stars in this cluster,
which are probably $\sim$ 3 Myr old do not show any emission feature. Thus these stars
do not posses any material around them which is left over from the parent
molecular cloud. 
                                                                                                                    
In this study we demonstrated that the low-resolution spectra of stars in the vicinity of a cluster
can be very useful to supplement the results obtained from photometry. We saved a lot of telescope
time by using the method
of slit-less spectroscopy, instead of obtaining spectra of individual stars.
This is a very effective method to study some of the characteristics
of stars in young clusters. Here, we used this method to identify the presence of Be stars and
confirm the photometrically identified PMS stars. This is the first time we used this method to identify
PMS stars in a cluster, we plan to extend this technique to more young clusters.

\acknowledgements
We thank the referee for encouraging comments. AS thanks A.E.Piskunov, Eswar Reddy and Maheswar for
helpful discussions.
This publication makes use of data products from the Two Micron All Sky Survey, 
which is a joint project of the University of Massachusetts and the Infrared 
Processing and Analysis Center/California Institute of Technology, funded by 
the National Aeronautics and Space Administration and the National Science Foundation.

\end{document}